\documentclass{article}
\usepackage{booktabs}
\usepackage{amsmath,graphicx,mlspconf}
\usepackage{amsfonts}
\usepackage{bm}
\usepackage{xcolor}

\usepackage{multirow} % in preamble
\usepackage{makecell} % in preamble

\usepackage[colorlinks=true,bookmarks=false]{hyperref}
\toappear{2025 IEEE International Workshop on Machine Learning for Signal Processing, Aug.\ 31-- Sep.\ 3, 2025, Istanbul, Turkey}

\title{Re-Bottleneck: Latent Re-Structuring for Neural Audio Autoencoders}
\name{%
   Dimitrios Bralios$^{1}$%
   \qquad Jonah Casebeer$^{2}$%
   \qquad Paris Smaragdis$^{1}$%\thanks{Thanks to XYZ agency for funding.}%
}
\address{%
   $^{1}$ University of Illinois Urbana-Champaign \quad%
   $^{2}$ Adobe Research%
}

\begin{document}
\ninept

\maketitle

\begin{abstract}
% Setting
Neural audio codecs and autoencoders have emerged as versatile models for audio compression, transmission, feature-extraction, and latent-space generation.
% Problem
However, a key limitation is that most are trained to maximize reconstruction fidelity, often neglecting the specific latent structure necessary for optimal performance in diverse downstream applications.
% Solution
We propose a simple, post-hoc framework to address this by modifying the bottleneck of a pre-trained autoencoder. Our method introduces a ``Re-Bottleneck", an inner bottleneck trained exclusively through latent space losses to instill user-defined structure.
% Experiments
We demonstrate the framework's effectiveness in three experiments.
First, we enforce an ordering on latent channels without sacrificing reconstruction quality. Second, we align latents with semantic embeddings, analyzing the impact on downstream diffusion modeling. Third, we introduce equivariance, ensuring that a filtering operation on the input waveform directly corresponds to a specific transformation in the latent space.
% Broader Message
Ultimately, our Re-Bottleneck framework offers a flexible and efficient way to tailor representations of neural audio models, enabling them to seamlessly meet the varied demands of different applications with minimal additional training.
\end{abstract}

%\newcommand{\cem}[1]{\textcolor{blue}{cem: #1}}
%
% Abstract
% Neural audio codecs and autoencoders have emerged as versatile models for audio compression, transmission, feature-extraction, and latent-space generation.
% However, a key limitation is that most are trained to maximize reconstruction fidelity, often neglecting the specific latent structure necessary for optimal performance in diverse downstream applications.
% We propose a simple, post-hoc framework to address this by modifying the bottleneck of a pre-trained autoencoder. Our method introduces a ``Re-Bottleneck", an inner bottleneck trained exclusively through latent space losses to instill user-defined structure.
% We demonstrate the framework's effectiveness in three experiments.
% First, we enforce an ordering on latent channels without sacrificing reconstruction quality. Second, we align latents with semantic embeddings, analyzing the impact on downstream diffusion modeling. Third, we introduce equivariance, ensuring that a filtering operation on the input waveform directly corresponds to a specific transformation in the latent space.
% Ultimately, our Re-Bottleneck framework offers a flexible and efficient way to tailor representations of neural audio models, enabling them to seamlessly meet the varied demands of different applications with minimal additional training.

\section{Introduction}

% Neural audio auto encoders are great, but there is a disconnect from the downstream applications
Neural audio autoencoders and codecs have become foundational in modern audio processing, enabling high-fidelity reconstruction from compact latent representations~\cite{zeghidour2021soundstream, defossez2023high, kumar2024high, evans2024stable}. These models are powerful components, driving innovation in areas like audio generation via next-token prediction~\cite{agostinelli2023musiclm, flores_garcia2023vampnet, borsos2023soundstorm}, latent diffusion~\cite{evans2024stable} and supporting tasks such as classification, enhancement, and source separation~\cite{mishra2024time, yip2024speech}. However, their primary training objective, optimizing reconstruction, often results in latent spaces that lack the specific structure required for optimal performance in many downstream applications. This creates a crucial disconnect: general-purpose latents from reconstruction-focused autoencoders are not inherently aligned with the needs of specialized audio tasks.

% People are creating new autoencoders to capture their desired properties
Addressing this structural deficit is challenging. Current practice typically involves either adapting downstream models to accommodate the autoencoder's arbitrary latent structure (e.g., requiring specialized token-prediction orders~\cite{agostinelli2023musiclm}, or coarse-to-fine diffusion~\cite{discodiff}), or, more commonly, undertaking the significant effort of redesigning and retraining the autoencoder from scratch to embed desired properties. Examples of this costly retraining include developing task-specific tokenizers for improved prediction~\cite{lemercier2024an} or designing models for semantic alignment directly in the latent space~\cite{zhangspeechtokenizer, kyutai2024moshi, wu2024codec}. While effective, these tailored autoencoders demand substantial architectural modifications and retraining on vast datasets, representing a major computational and development burden.

% Latent Diffusion --> Semantic allignment, equivariance diffuability
The value of explicitly structured latent spaces is particularly evident in generative models like latent diffusion, where structured representations have been shown to significantly improve training efficiency and generation quality in other domains~\cite{yao2025reconstruction, chen2025masked}. Applying similar principles to audio, such as incorporating semantic alignment or equivariance, holds great promise but would require modifying and retraining the core autoencoder. Given the increasing availability of powerful audio autoencoders pre-trained on large datasets, retraining from scratch to impose new structures is often impractical. Researchers are faced with the dilemma of using suboptimal existing latents or incurring the prohibitive cost of full autoencoder training.

% Problem training from scratch is costly, we have many open-source models
In this work, we ask: Can we efficiently impose desired structural properties onto the latent spaces of existing, off-the-shelf audio autoencoders without the prohibitive cost of full model retraining? To address this, we introduce \textbf{Re-Bottleneck}, a novel, lightweight framework. Re-Bottleneck is inspired by the Re-Encoder~\cite{bralios2025learning} and operates post-hoc by training a compact inner autoencoder, combined with an adversarial discriminator, within the latent bottleneck of a pre-trained model. This approach allows us to restructure the latent space using only latent-domain losses, circumventing complex waveform-level objectives and avoiding the extensive computation and tuning required for end-to-end autoencoder retraining. Re-Bottleneck offers a fast and flexible path to obtaining structured latents from pre-trained models, enabling their more effective use across diverse downstream audio tasks. We demonstrate Re-Bottleneck's capabilities through three distinct studies: %\cite{bralios2025learning}

\begin{itemize}
\item \textbf{Ordered Channels:} Enforcing monotonic ordering across latent channels to capture progressively finer detail, while maintaining near-full reconstruction fidelity.
\item \textbf{Semantic Alignment:} Aligning latent vectors with embeddings from pre-trained audio (e.g., BEATs \cite{chen2023beats}) and text (e.g., T5~\cite{raffel2020exploring}) models, and analyzing its impact on downstream diffusion-based audio generation.
\item \textbf{Equivariance Constraints:} Introducing transformation-equivariance in the latent domain, ensuring predictable latent transformations correspond to given filtering operations.
\end{itemize}

Collectively, these experiments highlight that lightweight, latent-only adaptations via the Re-Bottleneck framework can effectively yield structured and application-aligned representations without requiring modifications or retraining of the base autoencoder. Our framework thus provides an efficient, flexible, and rapid paradigm for tailoring the latent spaces of neural audio models, enabling researchers to seamlessly adapt powerful pre-trained models to meet the varied demands of different applications and supporting fast, iterative experimentation. % In the spirit of reproducible science and to enable use of this tool, we will open-source all checkpoints and training code after the anonymous review period.
In the spirit of reproducible science and to enable use of this tool, training code is available here\footnote{https://github.com/dbralios/rebottleneck}.

\section{Proposed Method}
\label{sec:method}

\subsection{Overview}

Our framework employs a pre-trained and frozen neural audio autoencoder, comprising an encoder $A_E$ and decoder $A_D$. The encoder transforms a potentially multi-channel input waveform $\bm{x} \in \mathbb{R}^{N \times L}$ to a compact latent representation $\bm{z} = A_E (\bm{x}) \in \mathbb{R}^{C \times T}$. The decoder reconstructs the waveform from the latent representation, $\hat{\bm{x}} = A_D(\bm{z})$. As is common in audio autoencoders, the autoencoder's bottleneck is parameterized as a variational autoencoder to facilitate learning within the latent space.

Operating exclusively within the latent domain of the base autoencoder, the \emph{Re-Encoder}($R$), consisting of an encoder $R_E$ and decoder $R_D$, maps the original latent representation $\bm{z}$ to an inner latent representation, the \emph{Re-Bottleneck}($\widetilde{\bm{z}}$):
\begin{equation}
\widetilde{\bm{z}} = R_E(\bm{z}) \in \mathbb{R}^{C^\prime \times T^\prime}
\end{equation}
where $C^\prime \times T^\prime$ are the dimensions of this inner latent space. The Re-Bottleneck, $\widetilde{\bm{z}}$, is specifically designed to exhibit desired properties relevant to the downstream task. For instance, these properties could include encouraging specific channels to represent distinct semantic features, enforcing a particular ordering of information, or promoting invariance/equivariance to certain input transformations.

The Re-Encoder then reconstructs an approximation of the original latent representation $\bm{z}$, denoted as $\hat{\bm{z}}$:
\begin{equation}
\quad \quad \hat{\bm{z}} =  R(\bm{z}) = R_D\bigl(R_E(\bm{z})\bigr).
\end{equation}
Crucially, only the Re-Encoder components ($R_E$ and $R_D$) are trained, while the pre-trained audio autoencoder ($A_E, A_D$) remains frozen. This strategy preserves the high-quality audio reconstruction capabilities of the original autoencoder while allowing targeted manipulation and structuring of the latent bottleneck $\bm{z}$ for the specific downstream task. This approach also sidesteps the significant challenge of training a high-quality audio decoder from scratch, which is known to be a notoriously difficult task, particularly in achieving perceptual fidelity. 

\subsection{Latent Space Training}
\label{sec:method_training}

Training the Re-Encoder $R$ primarily within the latent space offers significant advantages in terms of computational efficiency and training stability compared to methods that require computationally expensive forward and backward passes through the high-dimensional audio decoder $A_D$. Objectives defined in the waveform or spectral domains, often combined with discriminators operating on reconstructed audio, necessitate the materialization of waveforms and can introduce complex interactions with the structural regularization terms we apply to $\widetilde{\bm{z}}$. The fundamental objective for training the Re-Encoder is a latent space reconstruction loss, encouraging $\hat{\bm{z}}$ to be close to the original latent $\bm{z}$.
\begin{equation}
\mathcal{L}_{\mathrm{rec}}^R = \mathbb{E}_{\bm{x} \sim \mathcal{D}} \left[ \left\| A_E(\bm{x}) - R(A_E(\bm{x})) \right\|_2 \right],
\end{equation}
where $\mathcal{D}$ is the training dataset and $\bm{z}=A_E(\bm{x})$ serves as both the input to $R$ and the target for reconstruction. To ensure that the Re-Encoder's output latents $\hat{\bm{z}}$ remain within the distribution of latents produced by the base encoder $A_E$, we employ a single latent discriminator $D$. The role of $D$ is to distinguish  between ``real" latents (those directly from $A_E(\bm{x})$ and ``fake" latents (those reconstructed by the Re-Encoder, $R(A_E(\bm{x}))$. The discriminator's training objective is a standard adversarial loss:
\begin{equation}
\mathcal{L}^D_\mathrm{adv} = \mathbb{E}_{\bm{x} \sim \mathcal{D}} \left[ \left(1 - D\left(A_E(\bm{x})\right)\right)^2
+ 
D\left(R\left(A_E(\bm{x})\right)\right)^2
\right].
\end{equation}
The loss terms for the Re-Encoder $R$ include an adversarial term and a feature matching term, designed to encourage R to produce latents that fool the discriminator and match the feature statistics of real latents:
\begin{gather}
\mathcal{L}^{R}_{\mathrm{adv}} = \mathbb{E}_{\bm{x} \sim \mathcal{D}} \left[ \left(1 - D(R(A_E(\bm{x}))\right)^2
\right],\\
\mathcal{L}^{R}_\mathrm{fm} =  \mathbb{E}_{\bm{x} \sim \mathcal{D}} \left[\sum_{i=1}^N \frac{\| D^i (A_E(\bm{x})) - 
D^i(R(A_E(\bm{x})))\|_1}
{\|
D^i(R(A_E(\bm{x})))
\|_1}
\right],
\end{gather}
where $D^i$ denotes the feature map at the $i$-th layer of $D$.

Concurrently, the Re-Bottleneck $\widetilde{\bm{z}}$ can optionally be regularized with a KL-divergence term $\mathcal{L}_{\mathrm{kl}}^R$ to encourage its distribution towards a desired prior, similar to the base autoencoder's bottleneck regularization. Finally, and critically, we apply task-specific structural losses to the Re-Bottleneck $\widetilde{\bm{z}}$ . These losses, denoted collectively as $\mathcal{L}_{\mathrm{task}}^R$, are designed to enforce the desired properties on $\widetilde{\bm{z}}$ that are relevant to the downstream task (e.g., disentanglement losses, channel-wise sparsity, equivariance penalties). The specific form of $\mathcal{L}_{\mathrm{task}}^R$ depends on the particular downstream objective.

The total training objective for the Re-Encoder $R$ is a weighted sum of these loss components:
\begin{equation}
\mathcal{L}^R = \lambda_{\mathrm{rec}} \mathcal{L}_{\mathrm{rec}}^R + \lambda_{\mathrm{adv}} \mathcal{L}^{R}_{\mathrm{adv}} + \lambda_{\mathrm{fm}} \mathcal{L}^{R}_\mathrm{fm} + \lambda_{\mathrm{kl}} \mathcal{L}_{\mathrm{kl}}^R + \lambda_{\mathrm{task}} \mathcal{L}_{\mathrm{task}}^R,
\end{equation}
where the $\lambda$ are hyperparameters balancing the contribution of each loss term. Empirically, we find these weights easy to balance due to the well-behaved nature of training on VAE latents. The discriminator $D$ is trained using $\mathcal{L}_{\mathrm{adv}}^D$. 
To demonstrate Re-Bottleneck, we manipulate the base autoencoder's bottleneck in three distinct ways. 

\subsection{Re-Bottleneck Types}
\subsubsection{Ordered Re-Bottleneck}

We induce a monotonic, hierarchical ordering of the channels within the inner bottleneck, $\widetilde{\bm{z}}$, by implementing nested dropout during training~\cite{rippel2014learning}. At each training iteration, we randomly sample a prefix length $m \sim \mathcal{U}\{1, C^\prime\}$. This value, $m$ effectively sets the number of active channels in $\widetilde{\bm{z}}$. Based on the sampled $\widetilde{\bm{z}}$, we construct a binary mask $M^{(m)}$ of size $C^\prime \times T^\prime$, where $M^{(m)}_{c,t} = 1$ if the channel index $c$ is less than or equal to $m$ and 0 otherwise. This mask is then applied element-wise to the inner latent  representation $\widetilde{\bm{z}}$. Consequently, the forward pass through the rest of the model becomes
\begin{equation}
R^{(m)}(\bm{x}) = R_D \left( M^{(m)} \odot R_E(\bm{x})\right).
\end{equation}
We optimize $\mathcal{L}_{\mathrm{rec}}^R,\mathcal{L}^{R}_{\mathrm{adv}}, \mathcal{L}^{R}_\mathrm{fm}$ on $R^{(m)}(\bm{x})$, and sample an $m$ in every batch. This strategy forces $R$ to encode the most salient information into the earliest channels, thereby producing an ordered latent.

\subsubsection{Semantically-Aligned Re-Bottleneck}
\label{sec:sem_rebot}

We induce semantic structure in the inner bottleneck, $\widetilde{\bm{z}}$, by defining the task loss $\mathcal{L}_{\mathrm{task}}^R$ as a contrastive loss. This loss encourages similarity between $\widetilde{\bm{z}}$ and representations from a pretrained semantic encoder, $F_{\mathrm{sem}}$ (e.g. a self‐supervised audio model). Specifically, we first obtain a sequence of semantic embeddings $\bm{s} = F_{\mathrm{sem}}(\bm{x}) \in \mathbb{R}^{H \times K}$ from the input $\bm{x}$. These are then temporally-pooled to produce a semantic vector $\bar{\bm{s}} \in \mathbb{R}^H$. Similarly, we take the the inner latent representation $\widetilde{\bm{z}}$ apply a linear transform (linear probe) to map its $C^\prime$ dimensions to $H$ dimensions, and then temporally-pool to get  $\bar{\bm{z}} \in \mathbb{R}^H$. The constrastive loss is then calculated as:
\begin{equation}
\mathcal{L}_{\mathrm{task}}^R
= -\frac{1}{B}\sum_{i=1}^{B}
\log
\frac{\exp\bigl(\mathrm{sim}(\bar{\bm{z}}_i,\bar{\bm s}_i)/\tau\bigr)}
{\sum_{j=1}^{B}\exp\bigl(\mathrm{sim}(\bar{\bm{z}}_i,\bar{\bm s}_j)/\tau\bigr)}, 
\end{equation}
where \(\mathrm{sim}(u,v)=u^\top v/\|u\|\|v\|\), \(\tau\) is a temperature hyperparameter, and $B$ is the number of training samples in the batch. This objective encourages $R$ to capture the semantic content of the input in a straightforward and aligned manner within $\widetilde{\bm{z}}$.

\subsubsection{Equivariant Re‐Bottleneck}

We induce equivariance of the encoder $R_E$ with respect to a transformation $g(\cdot)$ by (with abuse of notation) training it such that $g(R_E(\bm{x}))$ is equal to $R_E(g(\bm{x}))$. This objective encourages $R_E$ to commute with $g(\cdot)$. Specifically, we consider a parametric Gaussian low-pass filter $g_{\alpha}(\cdot)$ with cutoff frequency $\alpha$, and its counterpart operating in the latent space, $h_{\alpha}(\cdot)$. The design of $h_{\alpha}(\cdot)$ mimics the behavior of a Gaussian filter in the frequency domain.  To achieve equivariance with respect to this filtering operation, we employ two strategies. First, an explicit loss:
\begin{equation}
\mathcal{L}_{\mathrm{task}}^R
= \mathbb{E}_{\bm x,\alpha} \left[
\left\|
  h_{\alpha}(\widetilde{\bm z})
  - 
  R_E(A_E(g_{\alpha}(\bm{x})))
\right\|_{2} \right].
\label{eq:explicit_loss}
\end{equation}
This loss explicitly trains the filtered latent representation $h_{\alpha}(\widetilde{\bm z})$ to match the latent representation of the filtered input $R_E(A_E(g_{\alpha}(\bm{x})))$. Inspired by \cite{kouzelis2025eq}, we modify our reconstruction term,
\begin{equation}
\mathcal{L}_{\mathrm{rec}}^R = \mathbb{E}_{\bm{x}, \alpha} \left[ \left\| A_E({g_{\alpha}(\bm x})) - R_D(h_\alpha(R_E(\bm z))) \right\|_2 \right],
\end{equation}
and apply analogous modifications to all discriminator terms.
This loss trains the full autoencoder $R$. Jointly, these losses train $R$ to commute with filtering, enforcing a unique regularization on $\widetilde{\bm{z}}$.

% Let
% $\mathcal{S}(\bm{x})\in\mathbb{R}^{F\times L^\prime}$ be the Short-Time Fourier Transform (STFT) of the input. We define
% $
% g_{\alpha}:\mathbb{R}^{F\times L^\prime}\to\mathbb{R}^{F\times L^\prime}
% $
% that denotes a parametric spectral filter (e.g. a low-pass filter with cutoff $\alpha$).  We then apply the filter
% \begin{equation}
% \bm{S} = \mathcal{S}(\bm{x}), \quad \bm{S}_{\alpha} = g_{\alpha}(\bm{S}), \quad \bm{x}_{\alpha} = \mathrm{iSTFT}(\bm{S}_{\alpha}),
% \end{equation}
% obtaining the filtered signal $\bm{x}_\alpha$ by performing the inverse STFT ($\mathrm{iSTFT}$). Both the original and filtered signals are encoded:
% $\bm{z} = A_E(\bm x)$ and $\bm z_{\alpha} = A_E(\bm x_{\alpha})$.

% Our goal is to enforce correspondence between the parametric spectral filter $g_\alpha$ and an inner latent‐space operator $h_\alpha :\mathbb{R}^{C^\prime\times T^\prime}\to\mathbb{R}^{C^\prime\times T^\prime}$ (e.g.\ channel‐wise scaling) that mirrors \(g_{\alpha}\).  Specifically, we require
% \begin{equation}
% \widetilde{\bm z}_{\alpha}
% \;\approx\;
% h_{\alpha}(\widetilde{\bm z}),
% \end{equation}
% where $\widetilde{\bm z}_{\alpha} = R_E(A_E(\bm x_\alpha))$.

\section{Experimental Setup}
\label{sec:exp_setup}

To validate the efficacy of Re-Bottlenecks, we conduct three experiments corresponding to the three Re-Bottleneck types above.  

\begin{figure*}[t!]
    \centering
    \includegraphics[width=\linewidth]{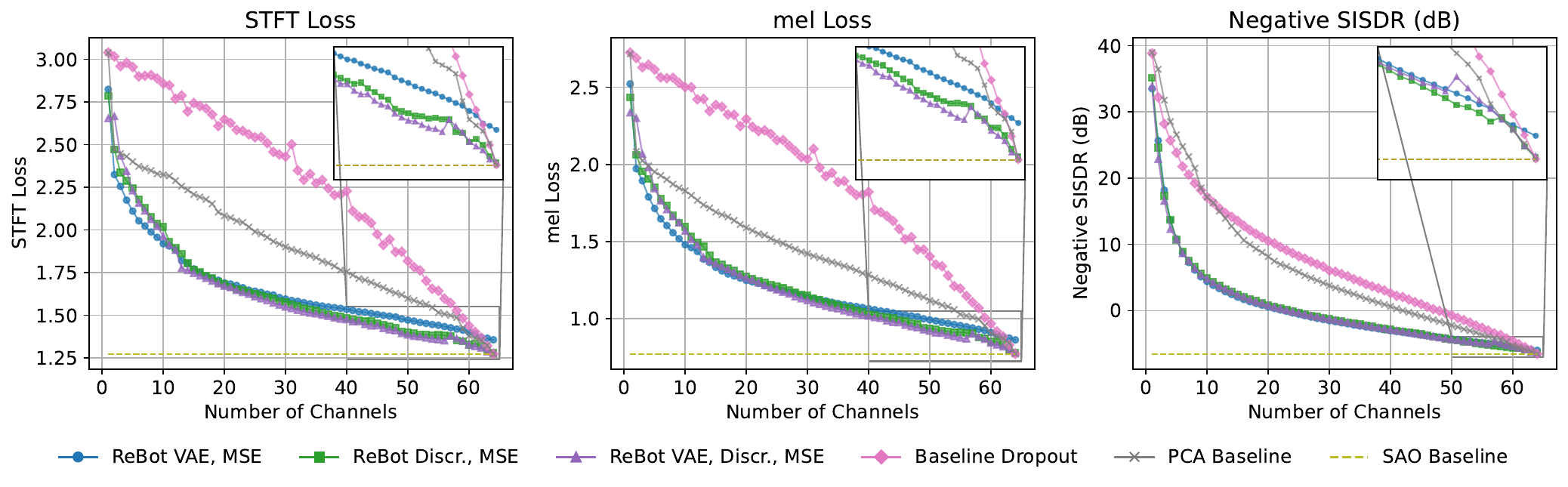}
    \caption{Reconstruction fidelity metrics (lower is better) as a function of the number of retained latent channels. We compare our Re-Bottlenecks against a PCA baseline (gray crosses), and a random dropout baseline (pink diamonds) which randomly selects channels to set to zero. The dashed horizontal line denotes the SAO full‐channel VAE baseline, and inset zooms highlight performance in the high‐channel regime.}

    \label{fig:ord_losses}
\end{figure*}

\begin{figure}[t]
    \centering
     \vspace{0pt}
        \centering
       \includegraphics[width=0.98 \columnwidth,
       trim=17.5pt 20pt 12.5pt 20pt]{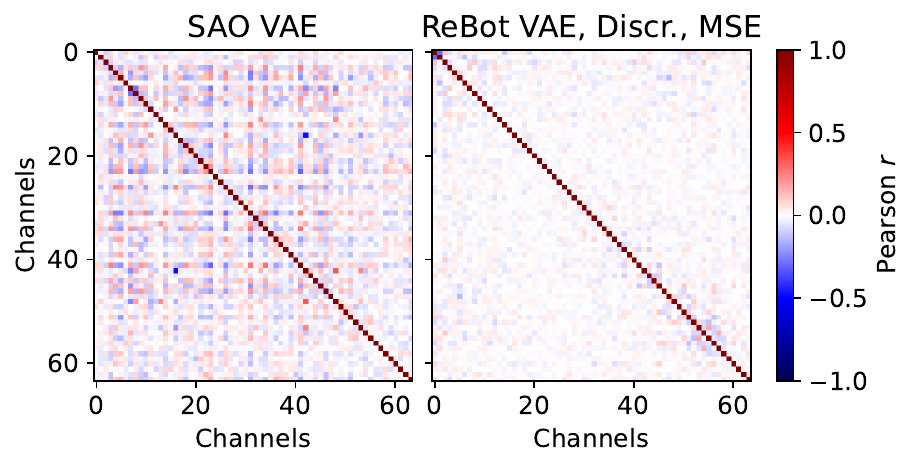}
        
\caption{Pearson correlation matrices of latent channels comparing the SAO VAE baseline (left) and our Re‐Bottleneck (right, VAE, discriminator, MSE). Lower off‐diagonal correlation in our model indicates a more decorrelated, less redundant latent representation.}

    \label{fig:ord_corr}
    % \vspace{-1 em}
\end{figure}

\begin{figure}[t]
    \centering
     \vspace{0pt}
        \centering
       \includegraphics[width=0.98 \columnwidth,
       trim=17.5pt 20pt 12.5pt 20pt]{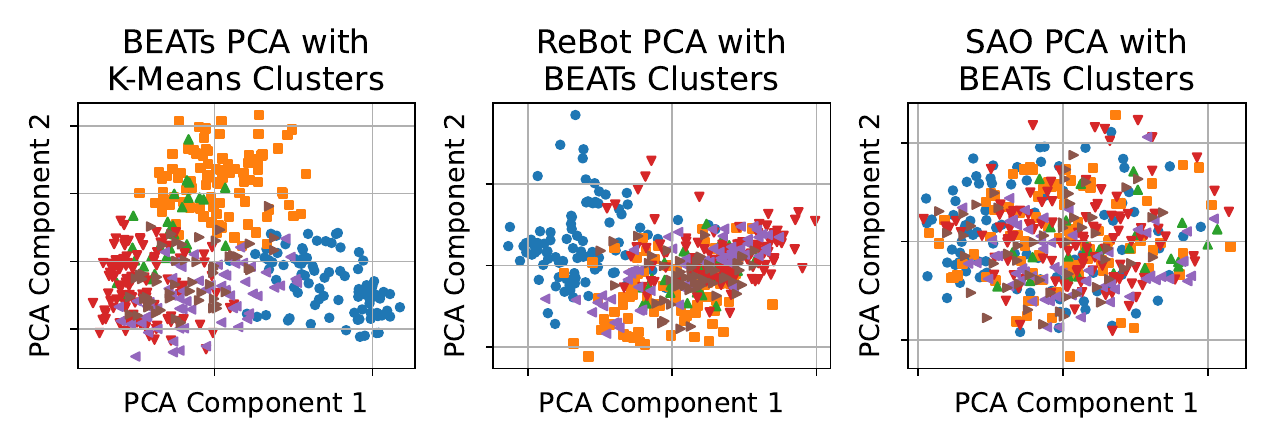}
       \caption{2D PCA visualization of mean‐pooled audio embeddings. Each point represents one audio file, colored by K‐means clusters computed in the BEATs embedding space. 
       %From left to right: (1) PCA of BEATs embeddings with their own cluster labels; (2) PCA of Re‐bottleneck embeddings using the same BEATs cluster assignments; (3) PCA of Stable Audio Open (SAO) embeddings with BEATs cluster labels. 
       This comparison highlights how semantic cluster structure is preserved or altered across different embedding spaces.}
    \label{fig:beats_kmeans}
    \vspace{-1.64em}
\end{figure}

\subsection{Re-Bottleneck Experiments}

% \subsubsection{Models \& Training}
For all experiments, we use the publicly released Stable Audio Open (SAO) VAE~\cite{evans2024stable} as our frozen $A$. This VAE compresses stereo $44.1$~KHz audio into $64$ channel latents at $21.5$~Hz. We use the provided checkpoint, originally trained for over 19 days on 32 GPUs. In the semantic alignment experiments we use \texttt{BEATs\_{iter3+}} \cite{chen2023beats} and $\texttt{T5-base}$ \cite{raffel2020exploring}.

The Re-Encoder model, comprising a symmetric encoder ($R_E$) and decoder ($R_D$) uses a ConvNeXt-V2 backbone~\cite{woo2023convnext}. Both encoder and decoder blocks are constructed from $4$ sequential ConvNeXt-V2 units. These units have hidden dimension $768$ and are placed between linear layers that match channel dimensions to the input and inner bottleneck. The complete model has $19.1$~M parameters. In most cases, the inner-bottleneck is parameterized as a VAE. The latent discriminator follows the architecture of the multi‐band discriminator~\cite{kumar2024high}. The discriminator takes the input inner latents, structured as a single-channel, 3-dimensional tensor. It is built from a sequence of Conv2d layers with 256 hidden channels, applying LeakyReLU activation after every layer except the last.

%Kernel sizes are set to $(3,7)$ except for the last layer where it is set to $(3,3)$. Stride and padding are both set to $(1,1)$. The number of hidden channels is $256$, while input and output channels are $1$.

% :
% \(
% \mathcal{L}^R
% = \lambda_{\mathrm{rec}}\,\mathcal{L}^R_{\mathrm{rec}}
% + \lambda_{\mathrm{KL}}\,\mathcal{L}^R_{\mathrm{KL}}
% + \lambda_{\mathrm{adv}}\,\mathcal{L}^R_{\mathrm{adv}}
% + \lambda_{\mathrm{fm}}\,\mathcal{L}^R_{\mathrm{fm}}\,,
% \)
% Training follows Sec.~\ref{sec:method_training}, optimizing a weighted sum of latent‐space objectives. 
For the base model variant, the loss weights are set as follows:
$\lambda_{\mathrm{rec}} = 1.0$, $\lambda_{\mathrm{kl}} = 10^{-4}$, $\lambda_{\mathrm{adv}} = 0.5$, and $\lambda_{\mathrm{fm}} = 1$. In the \emph{semantic} variant, $\lambda_{\mathrm{task}} = 2.5$, and in the \emph{equivariant} variant $\lambda_{\mathrm{task}} = 0.5$. General hyperparameters include optimizing the encoder–decoder with Adam using a learning rate of $5\times10^{-4}$ and the discriminator with $1\times10^{-4}$. The InfoNCE temperature $\tau$ was set to $0.07$. In the \emph{ordered} variant, the mask is only applied to 75\% of the batch. %Due to GPU memory limitations, we trained the semantic variant on 6~s chunks with a batch size of 32 for 12 epochs. For the ordered and equivariant variants, we halved the chunks and doubled the batch. 
We trained the ordered and equivariant variants on 3~s chunks with a batch size of 64 for 12 epochs. Due to GPU memory limitations, for the semantic variant we halved the batch size and doubled the chunk length.
We train on a single NVIDIA L40S GPU for $<48$~hours. 

\subsection{Diffusion Experiments}

We evaluate the downstream impact of Re-Bottleneck variants on text-to-audio diffusion by integrating them into the Stable Audio Open (SAO) pipeline \cite{evans2024stable}. Latent representations from all variants are first standardized to zero mean and unit variance based on training set statistics. We adopt the default SAO diffusion hyperparameters, omitting only weight decay. We train on 6-second chunks with a batch size of 72 across three NVIDIA L40S GPUs. Each variant is trained for 100K steps, requiring about four days per experiment.

\subsection{Data}

For training, we use the Jamendo-FMA-captions dataset~\cite{discodiff}, which comprises synthetically generated captions for the MTG-Jamendo~\cite{bogdanov2019mtg}
and FMA~\cite{defferrard2017fma} collections, totaling approximately 122K stereo files at 44.1 KHz ($\approx 8,000$~hours). We do not use Freesound due to being unable to locate a stereo version. For evaluation, we use the SongDescriber (no-vocals)~\cite{manco2023song, evans2024stable} benchmark derived from MTG-Jamendo. To prevent data leakage, the training split is sanitized by removing duplicate track IDs, exact audio-hash duplicates, and similar samples based on mel-spectrogram descriptors, following \cite{bralios2024generation}.

\subsection{Evaluation}

% We evaluate reconstruction fidelity using STFT, mel distance, and SISDR, all computed with the \texttt{auraloss} library~\cite{steinmetz2020auraloss}. In the case of the diffusion model, generation quality is measured by the FAD via the \texttt{fadtk} toolkit~\cite{gui2024adapting} configured with the CLAP-LAION-Music model. To assess prompt alignment, we report the CLAP score, which quantifies audio–text similarity~\cite{evans2024stable}. For SISDR and CLAP, higher is better. For STFT, mel distance, and FAD, lower is better. We measure representation alignment using Centered Kernel Alignment (CKA) \cite{kornblith2019similarity, cristianini2001kernel} and Projection Weighted Canonical Correlation Analysis (PWCCA)~\cite{morcos2018insights}, where higher is better.

We evaluate reconstruction fidelity using STFT and mel-spectrogram distances and SISDR (all computed with \texttt{auraloss}~\cite{steinmetz2020auraloss}). Generation quality for diffusion models is assessed with Fréchet Audio Distance (FAD) computed via \texttt{fadtk}~\cite{gui2024adapting} configured to use CLAP-LAION-Music embeddings. Prompt alignment is measured by the CLAP audio–text similarity score~\cite{evans2024stable}. Representation alignment is quantified with Centered Kernel Alignment (CKA)~\cite{kornblith2019similarity,cristianini2001kernel} and Projection-Weighted Canonical Correlation Analysis (PWCCA)~\cite{morcos2018insights}. Lower is better: STFT distance, mel distance, FAD. Higher is better: SISDR, CLAP, CKA, PWCCA.

% Finally, we measure generative diversity by computing precision and recall in the CLAP embedding space using $k$-nearest‐neighbors ($k=7$), following the protocol of~\cite{kynkaanniemi2019precision}.

\section{Experiments \& Results}
\label{sec:exp_results}
This section demonstrates the flexibility of the Re-Bottleneck framework by showcasing various latent space modifications. Training each Re-Bottleneck for these demonstrations required less than 48 GPU hours, representing under 0.33\% of the 14.5K GPU hours used for the base autoencoder.

\subsection{Ordered Re-Bottleneck}
In this experiment, we evaluate the ability of the Re-Bottleneck to impose a structured importance on latent channels and to induce decorrelation among them. We compared the reconstruction performance of a pre-trained SAO when latent channels were progressively removed, using different channel ordering strategies: random dropout, PCA, and three configurations of the Re-Bottleneck. As illustrated in Fig.~\ref{fig:ord_losses}, random dropout (pink) showed a rapid decline in performance as channels were removed. PCA (grey) provided a linear benchmark for performance degradation. The Re-Bottleneck trained with only an MSE loss (blue) achieved significantly better reconstruction performance than both random dropout and PCA across most channel counts. However, for the highest channel counts (i.e., when few channels were removed), random dropout and PCA slightly outperformed the MSE-only Re-Bottleneck. To improve performance, we introduced a latent discriminator loss, resulting in the Re-Bottleneck variant shown in green. This configuration improved performance over the MSE-only model, particularly for larger channel counts.

These first results demonstrate ordering but not orthogonality. To achieve this, we trained a Re-Bottleneck variant (shown in purple) using a combination of latent MSE, a latent discriminator, and a latent KL divergence loss. As demonstrated by the cross-channel correlation matrix in Fig~\ref{fig:ord_corr}, the original VAE latents exhibit significant off-diagonal correlation. In contrast, the latents produced by this Re-Bottleneck variant display a pronounced diagonal structure, indicating successful decorrelation. This experiment highlights the Re-Bottleneck framework's power in post-hoc tailoring of pre-trained latent spaces, demonstrating its ability to learn both an importance ordering and a decorrelated, potentially normalized, representation. This effectively positions the Re-Bottleneck as a learned, non-linear ``modern day PCA" for neural audio codecs.

\subsection{Semantically Aligned Re-Bottleneck}
We evaluate the ability of the Re-Bottleneck to align a pre-trained autoencoder latent space with that of a semantic model, while preserving invertibility. Unlike autoencoders focused on reconstruction, semantic models capture meaningful data attributes. We train the Re-Bottleneck using reconstruction, discriminator, and contrastive InfoNCE losses (aligning to BEATs ~\cite{chen2023beats}/T5~\cite{raffel2020exploring} embeddings) to instill semantic structure in the autoencoder latent space.

First, we visualize semantic structure via PCA of mean-pooled latents from the SongDescriber dataset, clustered by BEATs embeddings (Fig. \ref{fig:beats_kmeans}). The baseline VAE, trained for reconstruction, shows poor semantic clustering. The Re-Bottleneck trained with semantic alignment shows notably improved cluster separation compared to the VAE, while largely retaining reconstruction quality (Tab.~\ref{tab:recon_performance}).

% First, we investigate whether this training strategy successfully structures the geometry of our latent space to reflect the semantic organization captured by the semantic model. Fig. \ref{fig:beats_kmeans} presents PCA visualizations of different mean-pooled latent representations of the SongDescriber dataset, annotated with K-means clusters computed in the BEATs embedding space. The rightmost plot shows the results from our original pre-trained VAE. As anticipated, since this model was primarily trained for reconstruction and latent space whitening rather than semantic clustering, the BEATs clusters are largely indiscernible. %The leftmost plot shows the clustering results from the semantic model. Here, the clusters are clearly separated, demonstrating its effective capture of semantic distinctions, though it lacks a simple mechanism to reconstruct audio from its latent space. 
% The center plot shows the clustering for the pooled latents produced by the Re-Bottleneck model trained with the semantic alignment objective. Qualitatively, this model exhibits improved separation between clusters compared to the original VAE, while maintaining most of the base autoencoder's reconstruction quality as shown in Tab.~\ref{tab:recon_performance}.

% We measure alignment using Centered Kernel Alignment (CKA) \cite{kornblith2019similarity, cristianini2001kernel} and Projection Weighted Canonical Correlation Analysis (PWCCA)~\cite{morcos2018insights}  (Tab.~\ref{tab:alignment_metrics}). 
The baseline SAO bottleneck yielded 0.43 CKA / 0.63 PWCCA. A non-reconstructive upper bound (semantic loss only) achieved 0.69 CKA / 0.83 PWCCA. Our model including reconstruction loss alongside semantic objectives, achieved 0.70 CKA / 0.78 PWCCA. This shows that the model recovers nearly the upper bound's semantic alignment, indicating strong alignment and invertibility are largely compatible in the case of BEATs. Linear probe metrics (Sec.~\ref{sec:sem_rebot}) confirm that our linear probe effectively transfers semantic structure into the latent space. Reconstruction performance (Tab.~\ref{tab:recon_performance}) confirms invertibility. The semantically-aligned Re-Bottleneck showed ~5\% degradation versus the baseline autoencoder, trading this for a 20-60\% gain in semantic capture (CKA/PWCCA). This promising trade-off is achieved with efficient training ($<48$ GPU hours).

\begin{table}[t]
\centering
\caption{Reconstruction performance on 10~s chunks from SongDescriber (no vocals).}
\vspace{0.4 em}
\label{tab:recon_performance}
\resizebox{0.775 \columnwidth}{!}{
\begin{tabular}{lccc}
\toprule
Model               &   STFT ($\downarrow$) &    mel ($\downarrow$) &   SISDR ($\uparrow$)\\
\midrule
T5 aligned ReBot          & 1.29 & 0.79 & 6.2~dB\\
% \midrule
BEATs aligned ReBot             & 1.29 & 0.81 & 6.2~dB \\
 % + Input Masking     & 1.32 & 0.84 & 5.9~dB \\
\midrule
Equivariant ReBot             & 1.39 & 0.87 & 6.3~dB \\
\midrule
SAO Baseline       & 1.27 & 0.77 & 6.5~dB \\
\bottomrule
\end{tabular}
}
\end{table}

\begin{table}[t]
\centering
% \caption{Alignment Metrics (CKA / PWCCA)}

\caption{Alignment metrics (CKA ($\uparrow$) / PWCCA ($\uparrow$)) with mean-pooled T5 and BEATs embedding targets. ``B vs T" compares the mean-pooled bottleneck representation to the target, ``LP vs T" compares the linear probe to the target, and ``B vs LP" compares the bottleneck to the linear probe.}
\vspace{0.4em}
\label{tab:alignment_metrics}
\resizebox{\columnwidth}{!}{
\begin{tabular}{llccc}
\toprule
Model           & Target & B vs T        & LP vs T       & B vs LP        \\
\midrule
T5 aligned ReBot  & T5     & 0.19 / 0.71 & 0.18 / 0.73 & 0.98 / 1.00 \\          
\midrule
BEATs aligned ReBot    
                           & BEATs  & 0.70 / 0.78 & 0.65 / 0.82 & 0.96 / 0.98 \\
% {+ Input Masking} 
%                            & BEATs  & 0.69 / 0.77 & 0.67 / 0.81 & 0.96 / 0.98 \\
{- $\lambda_{\mathrm{rec}}, \lambda_{\mathrm{adv}}, \lambda_{\mathrm{fm}}=0$}
%{only $\mathcal{L}_\mathrm{sem}^R$}
                           & BEATs  & 0.69 / 0.83 & 0.66 / 0.83 & 0.97 / 1.00 \\
\midrule
SAO Baseline  & T5     & 0.15 / 0.68 & - & - \\
              & BEATs  & 0.43 / 0.63 & - & - \\
\bottomrule
\end{tabular}
}
\vspace{-1.2 em}
\end{table}

\subsection{Equivariant Re-Bottleneck}

\begin{table}[t]
\centering
\caption{(STFT ($\downarrow$) / mel ($\downarrow$)) distances to low-pass filtered target.} % for different cutoffs.}
\vspace{0.5 em}
\label{tab:distance_filtered_target}
\resizebox{\columnwidth}{!}{
\begin{tabular}{lccc}
\toprule
Cutoff & ReBot & AE Target   & SAO + Latent Filtering  \\
\midrule
1.4 KHz  & 0.98 / 0.91 & 0.88 / 0.78  & 5.08 / 3.76 \\
    % & mel  & 0.91 & 0.78 & 3.15 & 3.76 \\
2.8 KHz &  1.08 / 0.85 & 1.05 / 0.81  & 4.10 / 2.77 \\
    % & mel  & 0.85 & 0.81 & 2.07 & 2.77 \\
5.5 KHz &  1.17 / 0.81 & 1.16 / 0.79  & 2.60 / 1.81 \\
    % & mel  & 0.81 & 0.79 & 1.22 & 1.81 \\
\bottomrule
\vspace{0.3 em}
\end{tabular}
}
\end{table}

% \begin{table}[t]
% \centering
% \caption*{STFT ($\downarrow$) distance to low-pass filtered target.}
% \vspace{0.5 em}
% \label{tab:distance_filtered_target}

% \begin{tabular}{lccc}
% \toprule
% \makecell{Cutoff\\(kHz)} & \makecell{ReBot} & \makecell{AE Target} & \makecell{SAO + \\ Latent Filtering} \\
% \midrule
% 1.4 & 0.98 & 0.88 & 5.08 \\
% 2.8 & 1.08 & 1.05 & 4.10 \\
% 5.5 & 1.17 & 1.16 & 2.60 \\
% \bottomrule
% \end{tabular}

% \end{table}

% $\alpha$ & Metric & ReBot-Exp & ReBot & AE filtered vs filtered & AE full band vs filtered & SAO Reordered filtering \\
% \midrule
% 64  & STFT & 0.91 & 0.98 & 0.88 & 4.59 & 5.08 \\
%     & mel  & 0.89 & 0.91 & 0.78 & 3.15 & 3.76 \\
% 128 & STFT & 1.08 & 1.08 & 1.05 & 3.57 & 4.10 \\
%     & mel  & 0.84 & 0.85 & 0.81 & 2.07 & 2.77 \\
% 256 & STFT & 1.16 & 1.17 & 1.16 & 2.16 & 2.60 \\
%     & mel  & 0.80 & 0.81 & 0.79 & 1.22 & 1.81 \\

We evaluate the Re-Bottleneck on its ability to enable equivariance,  the property where a known transformation of the input audio corresponds to another known transform of its latent representation. Enforcing equivariance creates a structured latent space that reflects these input-domain changes. To demonstrate this, we apply corresponding filters: a Gaussian filter $G_\alpha[k] = \exp (-0.5\, k^2\,\alpha^{-2})$ in the STFT domain, and  $h_\alpha(c) = \exp (-0.5\, \mathrm{mel}(c)^{1.4}\,\alpha^{-1.4})$ in the latent space. This latent filter $h_\alpha$ is designed to mirror the STFT filter $G_\alpha$ by assuming latent channel index $c$ maps to a mel-scaled frequency, $\mathrm{mel}(c)$. During training, we sample the cutoff with a minimum value of $\alpha$ corresponding to $1.4$~KHz. 

Table~\ref{tab:distance_filtered_target} reports STFT and mel distances (``STFT / mel”) comparing the output of each method to the low-pass-filtered target audio at three cutoff frequencies. Our Re-Bottleneck applies the filter in its latent space before decoding. Its decoded output quality is comparable to that of the
filtered target autoencoded by the base SAO VAE (labeled ``AE Target" in the table), confirming effective filtering in the latent domain. In contrast, applying the same latent filter to a standard VAE's latent space (even after channel ordering attempts) yields substantially higher errors. Qualitatively, we verified that the spectral envelope of the Re-Bottleneck output matched the target.

%Finally, we do note the full-band reconstruction degradation in Tab.~\ref{fig:ord_losses} shows that the bottleneck re-structuring incurs a reconstruction penalty.

Next, we verify that the Re-Bottleneck learns a meaningfully different latent structure compared to the original VAE. In a toy experiment, we created 0~dB mixtures of clean audio and a chirp signal. We then attempted chirp removal by applying a magnitude-based mask to the latent representation before decoding. This resulted in 1.3~dB SISDR for the Re-Bottleneck, significantly outperforming the original VAE which yielded -1.3~dB SISDR with the same masking. This demonstrates Re-Bottleneck structures the latent space to better separate components, confirming the learned structural difference.

% Finally, we ablate the explicit loss (Eq.~\ref{eq:explicit_loss}) to confirm its contribution to learning the desired latent transformation. Training without it increases the explicit equivariance loss (Eq.~\ref{eq:explicit_loss}, i.e., the MSE between filtered latents in the STFT and latent domain) by over an order of magnitude compared to the model trained with the loss. This confirms the explicit loss is crucial for learning the desired latent equivariance property.

Finally, we ablate the explicit loss (Eq.~\ref{eq:explicit_loss}) to confirm its contribution to learning the desired latent transformation. Omitting this term causes the equivariance error to increase by more than an order of magnitude compared to the model trained with it. This confirms the explicit loss is crucial for learning the % desired %latent 
equivariance property.

% increases the explicit equivariance loss (Eq.~\ref{eq:explicit_loss}, i.e., the MSE between filtered latents in the STFT and latent domain) by over an order of magnitude compared to the model trained with the loss. This confirms the explicit loss is crucial for learning the desired latent equivariance property.

%Finally, we ablate the explicit loss (Eq.~\ref{eq:explicit_loss}) to confirm its contribution to learning the desired latent transformation. Training without this loss results in an explicit equivariance loss (Eq.~\ref{eq:explicit_loss}, measured as the MSE between filtered latents in the waveform and latent domain) over 10 times higher . This confirms the explicit loss is crucial for learning the desired latent equivariance property.

% # RENC wit explicit loss
% # Ideally Low
% # mse_E_tau_x-tau_E_x: 0.059829486538175776
% # We want this to be high
% # mse_E_tau_x-E_x: 0.10490028215614002
% # Is going to be high
% # mse_E_x-tau_E_x: 0.16512803510332588
% # RENC without explicit loss
% # Ideally Low
% # mse_E_tau_x-tau_E_x: 0.7369486895377326
% # We want this to be high
% # mse_E_tau_x-E_x: 0.48514675370714055
% # Is going to be high
% # mse_E_x-tau_E_x: 0.6654122396301261

\subsection{Investigating Diffusability}

\begin{figure}[t]
    \centering
     \vspace{0pt}
        \centering
       % \includegraphics[width=0.98 \columnwidth,
       % trim=17.5pt 20pt 12.5pt 20pt]{figs/clap_score_vs_step.pdf}
       % % \caption{One}
       \includegraphics[width=0.9 \columnwidth,
       trim=17.5pt 30pt 12.5pt 20pt]{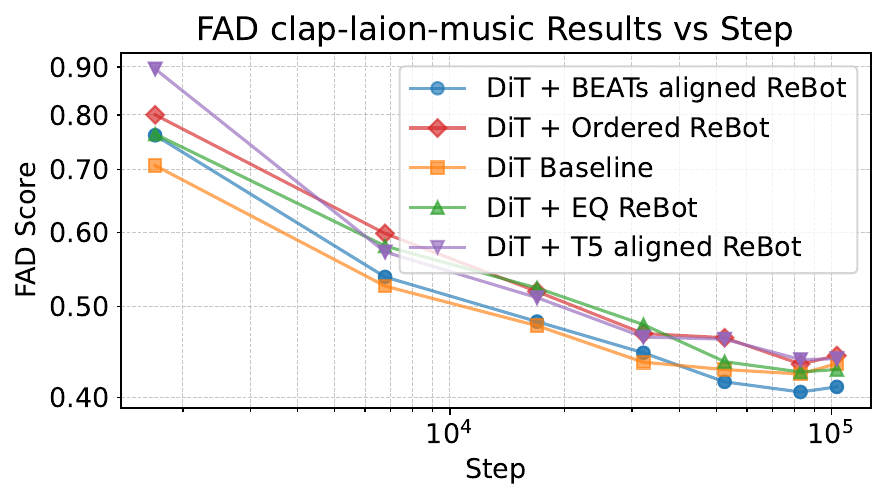}
       \caption{FAD scores ($\downarrow$) versus training steps, using the clap-laion-music backbone. The standard SAO (orange squares) starts best but is surpassed by the semantic Re-Bottleneck (blue circles). The ordered and T5-aligned Re-Bottlenecks do not improve downstream performance, while the equivariant matches baseline performance.}
    \label{fig:dit_scores}
    \vspace{-1.6 em}
\end{figure}
Leveraging Re-Bottlenecks to simplify latent space design, we evaluate their impact on downstream diffusion using the standard SAO pipeline. Fig.~\ref{fig:dit_scores} shows FAD scores over training iterations. After 100,000 steps, the baseline SAO scored 0.435 FAD and 0.185 CLAP Score. Our BEATs Re-Bottleneck improved performance to 0.411 FAD and 0.191 CLAP. The equivariant Re-Bottleneck (0.428 FAD, 0.176 CLAP) was closer to the baseline in terms of FAD. However, neither the T5-aligned (0.440 FAD, 0.181 CLAP) nor the ordered (0.443 FAD, 0.177 CLAP) variants showed improvement. This demonstrates the Re-Bottleneck's value as a versatile framework for latent prototyping.

% Metrics for DiT + BEATs aligned ReBot at epoch 60:
% 40     60               0.41058
% Metrics for DiT + BEATs aligned ReBot w/ Masking at epoch 60:
% 20     60              0.424316
% Metrics for DiT + Ordered ReBot at epoch 60:
% 20     60              0.443322
% Metrics for DiT Baseline at epoch 60:
% 20     60              0.434779
% Metrics for DiT + T5 aligned ReBot at epoch 60:
% 40     60              0.439926

% Metrics for DiT + T5 aligned ReBot at epoch 60:
% 40     60    0.181477

% Metrics for DiT + BEATs aligned ReBot at epoch 
% 40     60    0.190519

% Metrics for DiT + Ordered ReBot at epoch 60:
% 40     60    0.177163
% Metrics for DiT + BEATs aligned ReBot w/ 
% 40     60    0.186373
% Metrics for DiT Baseline at epoch 60:
% 40     60    0.184973

\section{Conclusion}
\label{sec:conclusion}
While pre-trained neural audio codecs excel at reconstruction, their default latent spaces often lack the specific structure needed for optimal performance across diverse downstream tasks. Our proposed Re-Bottleneck framework offers a flexible, post-hoc solution to this limitation. By applying targeted latent-space losses, demonstrated through experiments in achieving ordered channel importance, semantic alignment (while largely preserving reconstruction), and latent space equivariance, the Re-Bottleneck effectively instills user-defined structure into pre-trained representations. This approach is remarkably efficient, requiring minimal additional training—using less than a third of a percent of the compute needed for the original VAE. This efficiency allows neural audio models to readily and cost-effectively adapt to varying task demands without requiring expensive retraining of the base model.

\vfill
\pagebreak

\bibliographystyle{IEEEbib}
\bibliography{refs}
  
% \begingroup
%   \fontsize{9pt}{11pt}\selectfont
  
% \endgroup

\end{document}